# Linear processing of X-ray scattering patterns with missing pixels

Aliakbar Jafarpour

*Dept. of Biomolecular Mechanisms, Max-Planck Inst. for Medical Research, Jahnstr 29, 69120 Hiedelberg, Germany*
*jafarpour.a.j@ieee.org*

**Abstract:** X-ray scattering patterns from emerging single particle experiments have commonly many missing or contaminated pixels. This complicates different analyses including projections on Fourier or other basis functions (for noise suppression, compression, feature extraction, or retrieval of real-space patterns), as they require integration over all pixels. Here, we derive alternative formulations for *Discrete* Fourier Transform and a common orthogonal basis by explicit consideration of missing pixels and finite size. Such linear formulations exclude the nonlinear distortion that would be caused by multiplication of the complete scattering pattern with the mask function. Contrary to nonlinear and non-convex phase retrieval optimizations, such reduced-dimension formulations can be used to fully enforce the constraints and to retrieve unknown intensities in a linear fashion. Applications are demonstrated for some typical cases, and extensions to more general cases are discussed.

# 1. Introduction

## 1.1. Context of the problem

Typical formulations developed for the analysis of emerging single particle X-ray scattering patterns require [1] or imply [2,3,4,5] calculations with all pixel values. Real-life scattering patterns, however, have many missing or distorted pixel values. The effect of such pixels can be modeled as the product of the complete image and a binary mask, which is zero over bad pixels. This multiplication is a nonlinear operation and has a nontrivial signature in subsequent linear analyses (Fourier transform, low-pass filtering, projection on basis functions …).

Another complexity arises from the requirement of resampling a measured scattering pattern on a new grid, such as concentric [1] or eccentric [4] circular arcs, whereas the inherent grid of measured data is a rectangular array of pixels (for common small scattering angles). Analytic expression of measured intensities and calculating them on new grid points requires projections on a 2D basis. Projection would require integration over all pixels, unless special basis functions are employed that exclude bad pixels in their domain.

Within this context, phase-retrieval algorithms are used to retrieve not only the lost phase profile, but also the intensity profile of bad pixels. Despite existing potential for this additional task (due to redundancy of information), the additional output comes at the cost of making the final results even less reliable. Even in the ideal case (no noise, no missing pixels, and no truncation of the scattering pattern) phase retrieval algorithms perform a non-convex optimization with unknown level of success in approaching global optima. The additional burden of retrieving the intensities at bad pixels and/or suppressing the noise component of the known intensity makes this non-convex optimization (with reduced level of constraint) even more difficult.

## 1.2. Proposed approaches

Here we show that the two problems of 1) processing known intensities and 2) extraction of unknown intensities can be done independently of each other, independently of phase retrieval, and in a linear fashion. This can be done by utilizing the *a-priori* knowledge of the finite size of the object, yet without (hyper-) sensitivity to the specific shape or size of the boundaries [6]. Fig. 1 illustrates these linear schemes.

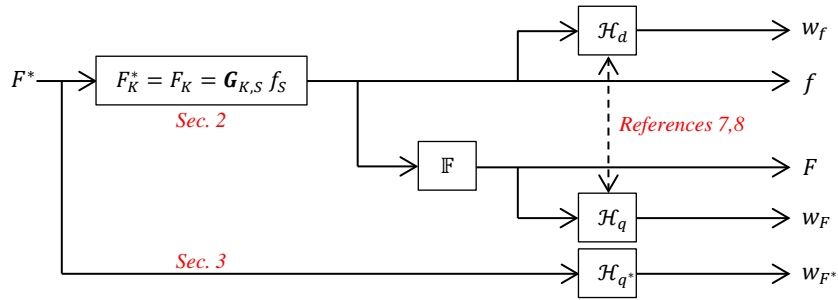

Fig. 1. Conceptual block-diagram illustrating the materials in subsequent Sections, namely the retrieval of missing intensities using linear regression (Sec. 2) and projection on limited-domain basis functions (Sec. 3). Projection on basis functions can be done in either the real- or the Fourier-space. The interrelation between expansions in these two spaces is addressed in the context of Zernike-Bessel expansions [7,8]. The block diagram $\mathbb{F}$ represents Fourier Transform, and the block diagrams $\mathcal{H}_q$, $\mathcal{H}_{q^*}$, and $\mathcal{H}_d$ represent arbitrary basis function expansions (Hilbert space vectors) in terms of different independent variables.

*1.3. Outline*

This report has been structured as follows. In Section 2, a simple reformulation of *Discrete* Fourier Transform and linear retrieval of missing intensities are addressed. Radial, angular, and the full 2D profiles of basis functions adapted to scattering patterns with missing pixels are addressed in Section 3, and brief conclusions are made in Section 4. Further discussions, mathematical derivations, and corresponding computer programs can be found in Appendices 1-7.

## 2. Projection on *conventional* basis functions with linear regression

*2.1. Reduced-dimension formulation of Discrete Fourier Transform (DFT)*

The inverse Fourier transform of an ideal measured scattering pattern (without taking square root) is the autocorrelation function of the projected density (Wiener-Khinchin theorem with the assumption of Geometrical model of scattering [9] and small scattering angles). This *2D Patterson profile* can be estimated without phase retrieval. Fig. 2 shows the common Fourier transform-based formulation and this alternative formulation of the scattered intensity (and the approach for retrieval of missing intensities).

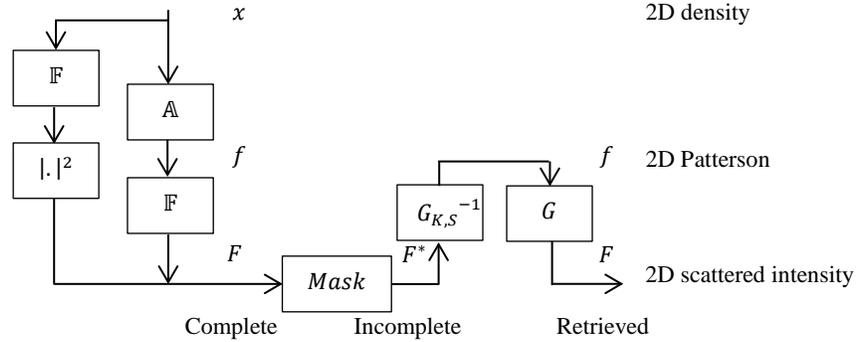

Fig. 2. Alternative formulations of the scattering pattern $F = \mathbb{F}\{\mathbb{A}[x]\} = |\mathbb{F}\{x\}|^2$, where the operators $\mathbb{A}$ and $\mathbb{F}$ represent *Continuous* Autocorrelation and Fourier Transform, respectively. The diagram shows interrelations between the directly-observable variable $F^*$ (scattered intensity with missing pixels), the sought variables $F$ (complete scattering pattern), and the Patterson profile $f$. The blocks $G$ and $G_{K,S}^{-1}$ represent the full- and reduced-dimension forms of *Discrete* Fourier Transform, and have been used here symbolically. Implicit (in-between the two blocks $G_{K,S}^{-1}$ and $G$) is wrapping ($2D \to 1D$) and zero-padding of the *nontrivial* Patterson profile to obtain the full 2D profile. The actual relation between these discrete Fourier operators and the conventional Fourier transform (the measured scattering pattern) has been shown in Fig. 3.

Given the linearity of 2D *Discrete* Fourier transform, the relation between a 2D function $f$ (Patterson profile) and its Fourier transform $F$ (scattering pattern), when rewritten as vectors, can be simply written as $F = \boldsymbol{G}f$, where $\boldsymbol{G}$ is the $N \times N$ *Kernel* matrix ($N$ is the number of pixels or data points in each 2D function). The elements of the Kernel matrix are complex exponentials on the unit circle in the complex plane, defined as $g_{(p,q),(r,s)} = \exp\left[-i\frac{2\pi}{N_{1D}}(pr + qs)\right]$, where $N_{1D}$ is the number of pixels *per coordinate* ($N_{1D}^2 = N$), and $0 \leq p, q, r, s < N_{1D}$. The integer pairs $(p, q)$ and $(r, s)$ serve as indices for the elements of $F$ and $f$, respectively.

Typically, $f$ is confined to a finite (support) region in the middle of the coordinate. This *Patterson support* (here simply referred to as support) is related to, yet different from the common "support region [of the density profile]").

We define the set $S$ to include all indices of the vector $f$ that lie in the support region (nontrivial values). So, $f$ can be split into a nontrivial component defined on $S$ and having $N_s$ values and another component comprising $N - N_s$ zeroes. The sets $K/U$ are also defined to include those indices of the vector $F$ that have known/unknown values, respectively. The measured intensity can also be split into two subsets $F_K/F_U$ with known/unknown pixel values. With these definitions, the sub-matrices $\boldsymbol{G}_{K,S}/\boldsymbol{G}_{U,S}$ include elements of the matrix $\boldsymbol{G}$ corresponding to the support region in $f$ and known/unknown values in $F$. Now, one can split the *Discrete* Fourier Transform equation $F = \boldsymbol{G}f$ into $F_K = \boldsymbol{G}_{K,S}f_S$ and $F_U = \boldsymbol{G}_{U,S}f_S$.

This reformulation has been exemplified in a 1D test case in Appendix 1. Applications and other considerations regarding this formulation have been detailed in Appendix 2. Matlab scripts for calculation and application of the $\boldsymbol{G}$ matrix have been presented in Appendix 7.

*2.2. Linear retrieval of missing intensities*

With estimated autocorrelation (Patterson) pattern using the linear regression $F_K = \boldsymbol{G}_{K,S}f_S$, unknown intensities can be then easily and directly calculated as $F_U = \boldsymbol{G}_{U,S}f_S$. With the missing intensities retrieved, a complete scattering pattern can be projected on conventional basis sets, or analyzed differently in a straightforward way.

Typically-large values of oversampling imply that $N_k \gg N_s$. So, one might choose a subset $L$ of $K$ and only some of the information as $F_L = \boldsymbol{G}_{L,S}f_S$ and solve for the sought pattern $f_S = \boldsymbol{G}_{L,S}^{-1}F_L$ (Note that $L \subset K$ and $N_s \leq N_L \leq N_K$). The sub-matrix $\boldsymbol{G}_{L,S}$ is data-independent and cannot be "ill-conditioned". It can be easily verified whether this sub-matrix is of full rank ($N_s$, corresponding to linearly-independent equations) or not, without knowing pixel intensities.

As an example, for a typical $1000 \times 1000$ scattering pattern ($N = 1000^2 = 10^6$) with $N_s = 100$ pixels in the support region, one needs the first $N_L$ pixels, where $N_s = 10^2 < N_L \sim 10^4 < N = 10^6$ to represent the full non-redundant information of the complete image. Smaller values of $N_L$ result in $G_{L,S}$ sub-matrices that are not of full rank ($Rank < N_s$). In practical cases with gaps in the indices of pixels, the minimum reliable number $N_L$ can be different and index-dependent.

The distinction between *Continuous* and *Discrete* Fourier Transforms is simple, yet crucial in our derivations. The linearity addressed to and utilized here as $F = \boldsymbol{G}f$ corresponds to *Discrete* Fourier Transform. The quadrants of an ideal scattering pattern (samples of a *Continuous* Fourier Transform on the discrete grid of a camera) should be first swapped to correspond to the employed discrete formulation, as shown in Fig. 3.

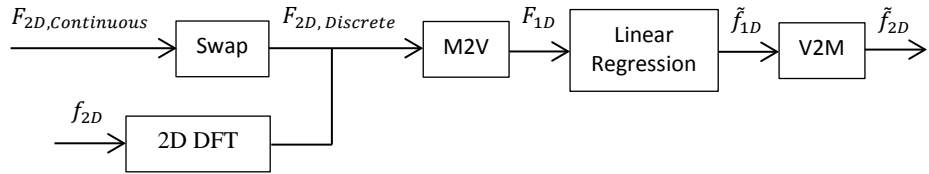

Fig. 3. Detailed block diagram showing the similarities and differences between *continuous* (measured) Fourier transform and the *discrete* (calculated) one. The M2V (Matrix to Vector) and the V2M (Vector to Matrix) conversions depend on the formulation and programming conventions. In a 1D formulation of an image (simple stack-based coding), the M2V and V2M may require no additional programming.

### 3. Basis functions excluding bad pixels

*3.1. Origin and modified form of basis functions*

Physical origins and typical topology/geometry of bad pixels and the need for alternative basis functions excluding bad pixels have been addressed in Appendix 3. Here we begin by tracing a common basis back to the differential equation of its origin.

At the core of Fourier transform are sinusoidal (complex exponential) functions. These are eigenfunctions of the ordinary differential equation $\psi'' + k^2\psi = 0$. The 2D equivalent is $\nabla^2\psi + k^2\psi = 0$, which is closely related to the simpler homogeneous Laplace equation $\nabla^2\psi = 0$. The solution of Laplace equation in the 2D Cartesian coordinate is simply split into two separable 1D ones; each satisfied by sinusoidal functions (of $x$ or $y$). In the polar coordinate, however, the angular variations are described by sinusoidal and radial variations by Bessel functions of the first and the second kind.

The sought polar-coordinate basis functions for a given geometry of good pixels and a given constraint on its boundary are eigenfunctions of the Laplace's operator as $\psi_{\nu,n}(r,\phi) = \Phi_\nu(\phi)R_{\nu,n}(r)$, where $\Phi_\nu(\phi) = [\alpha_\nu e^{i\nu\phi} + \beta_\nu e^{-i\nu\phi}]$ and $R_{\nu,n}(\phi) = [\gamma_{\nu,n}J_\nu(k_{\nu,n}r) + \eta_{\nu,n}Y_\nu(k_{\nu,n}r)]$. The non-trivial content of the coefficients $\{\alpha,\beta,\gamma,\eta\}$ is just two numbers (within a scale factor). These two numbers and the two *selection rules* for the choice of discrete $\{\nu, k_{\nu,n}\}$ values are determined by four boundary conditions; two on radial and two on angular functions. The scale factor has a magnitude determined using the extra convenient condition of orthonormality, and a constant complex phase factor chosen arbitrarily to have a simple formulation.

The most general case of specifying boundary conditions is specifying the ratio of the function and its (normal) derivative at the boundary, referred to as the Sturm-Liouville boundary condition for a 1D problem. Here we start with the assumption of 1) the geometry in the form of an annular ring limited to $a \leq r \leq b$ with 2) the soft boundary condition at $r = a$ (typical of diffraction patterns) and the hard boundary condition at $r = b$ (explicitly enforcing real-space *resolution*).

As an acoustic analogy, the sound generated by a drum can be decomposed into contributions from orthogonal resonance modes. The question is how to find the *modes of a torn drum* (loose at $r = a$ and anchored at $r = b$) and express a sound profile in terms of these new modes.

*3.2. Angular basis functions*

With full angular span, the continuity of $\Phi_\nu(\phi)$ at boundaries or $\Phi_\nu(0) = \Phi_\nu(2\pi)$ requires discrete values of $\nu$. A common choice for the weight coefficients $\alpha$ and $\beta$ in $\Phi_\nu(\phi) = [\alpha_\nu e^{i\nu\phi} + \beta_\nu e^{-i\nu\phi}]$ is the orthonormal basis $\Phi_\nu(\phi) = \frac{1}{\sqrt{2\pi}} e^{i\nu\phi}$. Physically, it represents a single azimuthally-propagating wave without reflections. Alternative choices of $\alpha_\nu$ and $\beta_\nu$ will be addressed below and also in Section 3.6.

The case of incomplete angular span requires specific assumption about (or tweaking a parameter corresponding to) boundary conditions. In some cases, a reasonable assumption can be smooth angular variations and hence $\Phi'(\phi) = 0$ at boundaries (Note that even with smooth angular variations $\partial I/\partial \phi \sim 0$, one can still expect considerable slopes of the radial component and hence the entire function $dI = (\partial I/\partial r)dr + (\partial I/\partial \phi)d\phi$ at angular boundaries).

We consider an annular sector centered at $\phi_0$ with a span of $\Delta\phi$ as $|\phi - \phi_0| \leq \Delta\phi/2$ to be the domain of definition. Enforcing soft boundary condition at both ends means $\Phi'_\nu(\phi) = i\nu[\alpha_\nu e^{i\nu\phi} - \beta_\nu e^{-i\nu\phi}]_{\phi=\phi_0\pm\Delta\phi/2} = 0$. Nontrivial values of $\alpha_\nu$ and $\beta_\nu$ determine the selection rule $\nu = m(\pi/\Delta\phi)$, where $m$ is an integer. The dependence of $\alpha_\nu$ and $\beta_\nu$ on each other and the requirement of orthonormality (and a convenient choice of the complex phase factor) determine the final form of the angular function as

$$\Phi_\nu(\phi) = \left[\frac{\Delta\phi}{2}(1+\delta_{\nu,0})\right]^{-1/2} cos[\nu(\phi-\phi_0-\Delta\phi/2)]$$

The parameter $\nu(\Delta\phi/\pi)$ assumes all integers; $\Phi_{-\nu}(\phi) = \Phi_\nu(\phi)$; and $\Phi_{\nu=0}(\phi) = 1/\sqrt{\Delta\phi}$. A complete set of *independent* basis functions corresponds to $\nu(\Delta\phi/\pi) \in \{0,1,2,3\ldots\}$.

Physically, the effect of limiting the angular span is converting the uni-directional flow of the single azimuthal wave $e^{i\nu\phi}$ to a standing wave pattern with anti-nodes at soft boundaries.

For $\Delta\phi = \pi/N$ ($N$ a natural number), the selection rule implies that the angular modes are not only discrete, but also limited to the harmonics $\{..., -3N, -2N, -N, 0, N, 2N, 3N, ...\}$. For other values of $\Delta\phi$ (for instance $\Delta\phi = \pi/2.5$), the discrete angular spectrum is preserved. However, non-integer values of $\nu$ will have different orthogonality relations of radial components (compared to the relations in the case of integer $\nu$'s). For simplicity, we will limit the discussion to the case of $\Delta\phi = \pi/N$ only.

### 3.3. Radial basis functions

The general Sturm Liouville boundary condition for the function $R_{\nu,n}(r)$ at radial positions $r = a$ and $r = b$ can be written as

$$R(a)\cos(c_1) - aR'(a)\sin(c_1) = 0$$
$$R(b)\cos(c_2) - bR'(b)\sin(c_2) = 0$$

where $c_1$ and $c_2$ are two constants. With the assumption of a soft limit at $r = a$ and a hard limit at $r = b$, these constants can be selected as $c_1 = \pi/2$ and $c_2 = 0$, or $aR'(a) = R(b) = 0$. As such, the Sturm Liouville for the function $R(r) = AJ_\nu(kr) + BY_\nu(kr)$ at the two ends $a$ and $b$ can be rewritten as

$$AkJ'_\nu(ka) + BkY'_\nu(ka) = 0$$
$$AJ_\nu(kb) + BY_\nu(kb) = 0$$

By deleting the ratio $B/A$, a transcendental equation in terms of $k$ is obtained. This *dispersion equation* characterizes the discrete allowed modes $k = k_{\nu,n}$ (for a given order $\nu$ and radii $a$ and $b$), and can be written as

$$J_\nu(kb)/Y_\nu(kb) + J'_\nu(ka)/Y'_\nu(ka) = 0$$

Some useful properties of Bessel functions and also Matlab scripts for solving the dispersion equation can be found in Appendices 5 and 7, respectively. Once the discrete modes $k = k_{\nu,n}(a, b, \nu, n)$ are known, the ratio $B/A$ and hence the function $R$ (within a scale factor $A$) are known:

$$R_{\nu,n}(r) = J_\nu(k_{\nu,n}r) - \frac{J_\nu(k_{\nu,n}b)}{Y_\nu(k_{\nu,n}b)}Y_\nu(k_{\nu,n}r)$$

A convenient choice for the scale factor $A$ is one that makes $R_{\nu,n}$ orthonormal. While this normalization can be done computationally, it is also formulated analytically in Appendix 6. The overlap integral, expressed in terms of the auxiliary function $\phi$ (defined in the same Appendix), is

$$\langle R_{\nu,n}|R_{\nu,n}\rangle = \phi_n^{J,J}(\nu,b,a) - 2\left[\frac{J_\nu(k_{\nu,n}b)}{Y_\nu(k_{\nu,n}b)}\right]\phi_n^{J,Y}(\nu,b,a) + \left[\frac{J_\nu(k_{\nu,n}b)}{Y_\nu(k_{\nu,n}b)}\right]^2 \phi_n^{Y,Y}(\nu,b,a)$$

As such, the orthonormal radial basis functions (for allowed integer values of $\nu$) can be written as

$$R_{\nu,n}^{norm} = R_{\nu,n}/\langle R_{\nu,n}|R_{\nu,n}\rangle^{1/2}$$

### 3.4. Full 2D basis functions

The full 2D orthonormal basis functions will be $R_{\nu,n}^{norm}(r)\Phi_\nu^{norm}(\phi)$, as shown for the first few orders in Fig. 4. Note that Bessel functions of the first and the second kind both satisfy the relation $B_{-\nu}(x) = (-1)^\nu B_\nu(x)$. As such, the same property holds for the radial functions $R_{-\nu,n} = (-1)^\nu R_{\nu,n}$. A similar relation also holds for the angular functions $\Phi_{-\nu} = (-1)^\nu \Phi_\nu$ for full angular span and $\Phi_{-\nu} = \Phi_\nu$ for a limited one.

Linear combinations of 2D basis functions can also define new 2D basis functions. Specifically, $R_{\nu,n}\Phi_\nu$ and $R_{-\nu,n}\Phi_{-\nu}$ may be combined to generate $R_{\nu,n}^{even}(r)\cos(\nu\phi)$ and $R_{\nu,n}^{odd}(r)\sin(\nu\phi)$ with $\nu \geq 0$, as in the definition of Zernike polynomials for real functions

[7]. However, even for real functions, it may be helpful to use complex functions Φ for specific pattern recognition and rotation-invariant feature extraction purposes. This is the basis of the so-called Pseudo-Zernike basis set and relevant to the MPEG-7 standard for multimedia content description [10].

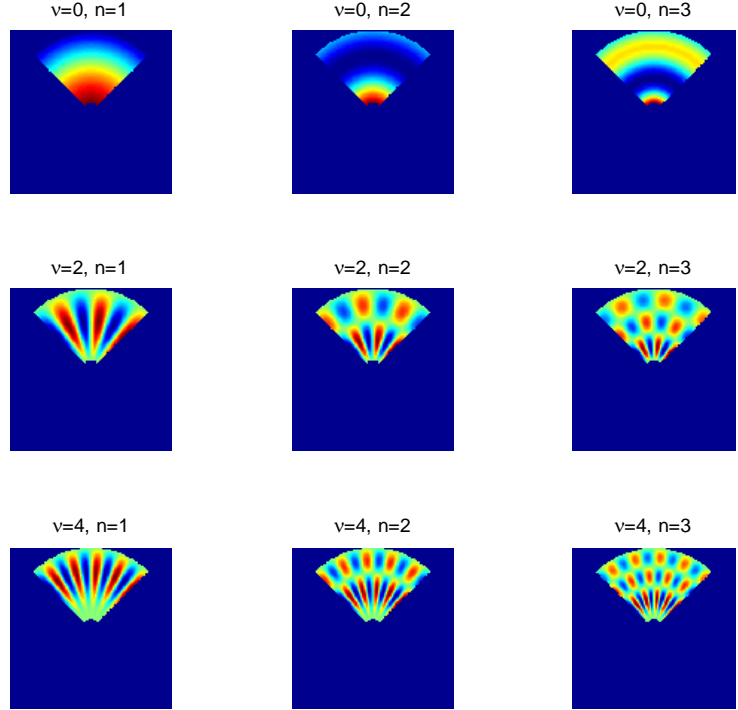

Fig. 4. The first few limited-domain basis functions. The 2D patterns along a row (column) have the same angular (radial) index. The angular span is $\pi/2$, and the radial limits are $a = 0.1R_{Max}$ and $b = 0.9R_{Max}$. Analytical basis functions have been calculated on a $200 \times 200$ discrete grid.

### 3.5. Projections on and filtering with basis functions

If a function $f$ can be expanded in terms of an orthonormal basis $\{\psi_{v,n}\}$; i.e., $f(\mathbf{r}) = \sum_{v,n} c_{v,n} \psi_{v,n}$, then the weight factors can be written as

$$c_{v,n} = \langle f(\mathbf{r}) | \psi_{v,n} \rangle = \int_{2\pi} d\phi \int_{r=a}^{r=b} f(\mathbf{r}) \psi_{v,n}^*(\mathbf{r}) r dr$$

By rewriting the above double integral in Cartesian coordinate (and changing the differential area from $rd\phi dr$ to $dxdy$), the numerical calculation can be done on a uniformly-sampled Cartesian grid (measured data). Discretization issues and uncertainties associated with resampling a Cartesian grid (measured data) on a circular grid affect only the pixels on the boundaries (contrary to a numerical approach, in which all rings require this resampling, and different techniques for "small" and "large" values of radius should be employed [4]). The rings at the boundaries have also the interesting property of being locally nearly-uniform or nearly-smooth.

Filtering with basis functions is paramount to limiting the number and indices of basis functions. Excluding higher-orders is low-pass filtering *in the discrete spectrum of the given*

*basis set*. Limiting angular variations can suppress distortions caused by strong total external *reflections* contaminating weak *scattering* patterns [7,11].

A common useful analysis of X-ray scattering patterns (of either nano-crystals or single particles) is angular averaging (virtual powder pattern). With basis functions, such 1D analyses require projecting data only on $\psi_{\nu=0,n}$ functions, as exemplified in Fig. 5. This filtering scheme can be implemented irrespective of specific selection rules for $\nu$ and specific forms of angular basis functions $\Phi_{\nu \neq 0}(\phi)$.

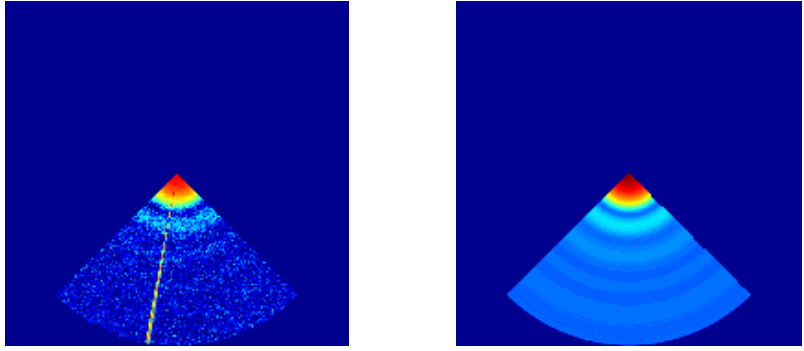

Fig. 5. Scattering pattern of a homogeneous sphere: (Left) The pattern is not only defined on a limited domain, but also contaminated with random noise and an angularly-localized (streak) component. (Right) Retrieval of the original (ideal) pattern by projecting the data on the limited-domain basis functions without angular variations. The angular span is 90 degrees, and the smaller radius $a$ is reduced to a small (sub-pixel, yet nonzero) value for improved numerical accuracy. The jet streak has been modeled with a Gaussian function of the angular coordinate, and the noise is a scaled zero-mean ensemble of normal distribution. The amplitudes of the noise and the streak have been tweaked independently to have tangible distortions, while preserving the strongest sought features. While the filtered pattern has always spherical symmetry, its resemblance to the target scattering pattern is affected by the type and strength of distortion.

### 3.6. Parity and further considerations regarding limited-domain basis sets

An important pre-processing step can be the decomposition of an image into even and odd angular components as $f(r,\phi) = f_{even}(r,\phi) + f_{odd}(r,\phi)$, where $f_{even}(r,\phi) = [f(r,\phi) + f(r, 2\phi_0 - \phi)]/2$ and $f_{odd}(r,\phi) = [f(r,\phi) - f(r, 2\phi_0 - \phi)]/2$. Similar to the case of periodic 1D functions, a basis comprising even functions (cosine-like) and a basis comprising odd functions (sine-like) may be both necessary for a complete description (Fourier-series-like) with alternative or unknown boundary conditions.

For a given image and with unknown angular boundary conditions, the weights of the counter-propagating azimuthal waves ($\alpha_\nu$ and $\beta_\nu$ in $\Phi_\nu(\phi) = [\alpha_\nu e^{i\nu\phi} + \beta_\nu e^{-i\nu\phi}]$) can be chosen by the best fit to experimental data ("practical completeness"). Note that the flexibility (or ambiguity) regarding boundary conditions affects not only the functional form of angular basis functions $\Phi_\nu(\phi)$, but also the selection rule for $\nu$ in the first place. Table 1 lists some different possibilities with soft/hard boundary conditions on the two boundaries. An arbitrary linear combination of all these functions (at a given radius) is simply a Fourier series in terms of $\phi$ with the fundamental angular frequency of $\nu_0 = \pi/(2\Delta\phi)$.

**Table 1: Functional forms and selection rules of angular basis functions with different boundary conditions**

| $\phi_0 + \Delta\phi/2$ | $\phi_0 - \Delta\phi/2$ | $\Phi_\nu(\phi)$ | $\nu$ |
|---|---|---|---|
| Soft | Soft | $cos[\nu(\phi - \phi_0 - \Delta\phi/2)]$ | $m(\pi/\Delta\phi)$ |
| Soft | Hard | $cos[\nu(\phi - \phi_0 - \Delta\phi/2)]$ | $(m + 1/2)(\pi/\Delta\phi)$ |
| Hard | Soft | $sin[\nu(\phi - \phi_0 - \Delta\phi/2)]$ | $(m + 1/2)(\pi/\Delta\phi)$ |
| Hard | Hard | $sin[\nu(\phi - \phi_0 - \Delta\phi/2)]$ | $m(\pi/\Delta\phi)$ |

For feature extraction, completeness can be less crucial (nearly-similar images still possess nearly-similar projections). However, for filtering or reconstruction purposes, one should consider the possibility and significance of an incomplete basis.

*3.7. Further considerations regarding limited-domain basis sets*

Further details regarding alternative (convex or concave) borders of the domain of definition and curved domains (spherical caps) have been addressed in Appendix 4.

**4. Concluding remarks**

The rich arsenal of linear system theory can be used in handling X-ray scattering patterns with missing pixels for noise suppression, compression, feature extraction, or retrieval of real-space patterns. It can be done either directly in the form of matrix multiplication or indirectly in the form of linear regression. These linear formulations are done by a reduced-dimension reformulation of *Discrete* Fourier Transform or enforcing explicit boundary conditions on (sub-) domains of known pixels. The rank of the regression (sub-) matrix and the completeness of the basis functions are important aspects of these formulations.

# Appendix 1: Example of reduced-dimension *Discrete* Fourier Transform

The reduced-dimension formulation is illustrated in Fig. 6 for the simpler case of a 1D function. The real-space profile $f(r)$ has been defined over a 7-point domain with a 3-element support (with nontrivial values) in the middle. The Fourier-space function $F(p)$ has 3 missing pixels. The core of this reformulation is reducing the Kernel matrix $\boldsymbol{G}$ (in this case with a size of $7 \times 7$) to the appropriate sub-matrix $\boldsymbol{G}_{K,S}$ (in this case with a size of $4 \times 3$) that would define a linear regression problem.

Vector and matrix indices are consistent and start from zero (not one). Indexing the Kernel matrix elements may be done with a single variable $l$ or two variables $(p, r)$, where $l = rN + p$, $0 \leq p, r < N$, and $0 \leq l < N^2$ (in this case $N = 7$). With this convention for indices, the Kernel matrix elements are simply $\boldsymbol{G}_{p,r} = \boldsymbol{G}_{l(p,r)} = \exp\left(-i\frac{2\pi}{N}pr\right)$.

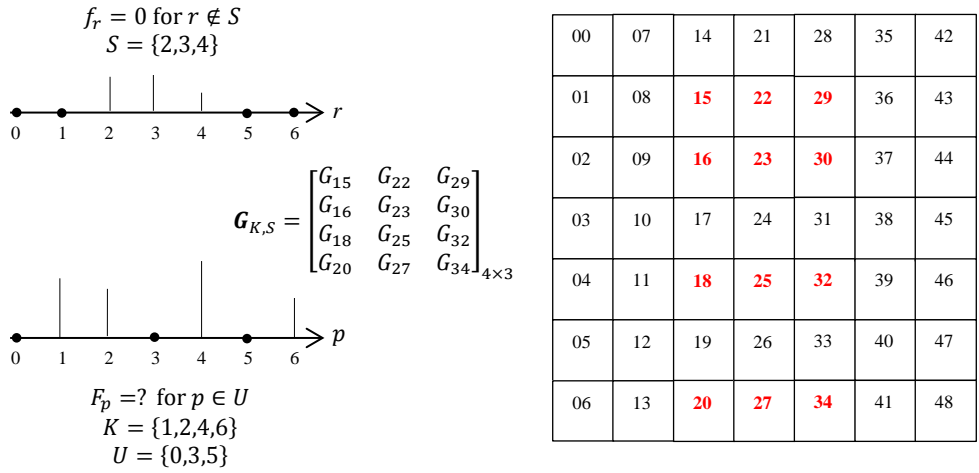

**Fig.** 6 Kernel matrix decomposition into sub-matrices corresponding to pixel subsets in real-space and Fourier space

# Appendix 2: Further aspects of reduced-dimension *Discrete* Fourier Transform

## A2.1. Numerical and electromagnetic benchmarking

The over-determined linear regression can also be used for benchmarking purposes, by splitting the data in two halves, for instance, and comparing the estimations. It can also be used by deliberately excluding very low pixel values and/or pixel values likely to have been contaminated. A measured pattern may include reflections (in addition to scattering) or a non-trivial offset. Even pure scattering may not follow the simple Fourier transform model [9]. Trying different subsets of $F_K$ in these linear regressions and comparing the estimated values of $f_S$ can be a good indicator of if the pattern can indeed correspond to a Fourier transform. Such results, even if not completely conclusive, can be more meaningful compared to those obtained by phase retrieval.

## A2.2. Fit with linear regression to other basis functions

Once the unknown intensities are retrieved, projection on conventional basis functions (coordinate axes of the Hilbert space $\mathcal{H}_q$ in Fig. 1) is straightforward. Also with known

Patterson profile (estimated without mask-induced distortions), projections on the coordinates of the Hilbert space $\mathcal{H}_d$ are straightforward. A special appropriate choice uses Bessel functions for $\mathcal{H}_q$ and Zernike polynomials for $\mathcal{H}_d$ [7,8].

A basis set is characterized with a Kernel function $g_{(p,q),(r,s)}$ and hence a Kernel matrix $\boldsymbol{G}_1$. In a practical discrete problem, numerical discretization errors may be suppressed if the two operations of inverse Fourier transform $f = \mathbb{F}^{-1}\{F\}$ and projecting the Patterson profile $f$ onto a basis with Kernel matrix $\boldsymbol{G}_1$ can be combined into a single linear regression. It replaces the Kernel matrix of the *Discrete* Fourier transform $\boldsymbol{G}$ with $\boldsymbol{G}\boldsymbol{G}_1^{-1}$.

*A2.3. Sensitivity to the exact boundary of Support*

Contrary to the case of phase retrieval [6], a rough estimation of the support region is not critical. A support slightly larger than the actual one is fine, and only sacrifices the level of being over-determined. In a noise-free problem, it introduces no error. Also, the geometry and even the topology of the known and unknown pixels are not that important (bad pixels can be distributed as multiple disconnected regions with arbitrary boundaries).

*A2.4. Extension to 3D and phase retrieval cases*

Linear retrieval of missing intensities can also be applied to the 3D case, in which the 3D Fourier transform is constructed by orienting and merging different measured 2D snapshots. Uncertainties associated with noise, estimation of orientations, and interpolation errors [2,3,4,5] can be suppressed using the over-determined linear regression problem.

With the missing pixels recovered, conventional 3D phase retrieval algorithms are expected to be more successful thanks to more constraints. Furthermore, the problem of phase retrieval itself can also be formulated in a similar lower-dimensional form. While the resulting regression is nonlinear, the optimization has now a significantly-reduced (*curse of*) dimensionality and also fully-enforced constraints.

**Appendix 3: On bad pixels and the need for basis sets excluding them**

*A3.1. Topology and physical origin of bad pixels*

Linear regression with the Kernel matrix, as exemplified for the Fourier operator in Section 2, is a discrete problem independent of the distribution of bad pixels. For defining basis functions limited to bad pixels, however, explicit knowledge and appropriate formulation of their geometry (shape of the boundary) is required. Physically, bad pixels fall into different categories, including
- A small closed region in the middle (coincident with a beam-stop)
- A large open region in the middle (in-between two isolated cameras; for efficient and tunable sampling of scattering angles)
- The shadow of front camera on the back camera
- Scattering and/or reflections from background medium (liquid jet carrying the object of interest)
- Optoelectronic and /or electronic distortions (a pixel being damaged, saturated, or contaminated with the charge "spilled" out of quantum wells)

*A3.2. Need for limited-domain basis functions*

With the missing intensities recovered, one can project scattering patterns on any conventional basis set. However, this retrieval is based on an ideal model of the scattering pattern. In practice,
- While linear regression has generally more robustness compared to a nonlinear and non-convex optimization (phase retrieval), the level of this robustness is not clear and can be lower than that offered by orthogonal projections.

- The retrieval of missing intensities discards authentic yet undesired optical intensity (such as grazing incidence *reflection* in addition to *scattering*) that make the use of Fourier Kernel questionable [9].
- Analog processing of transduced optical intensity can give rise to negative pixel values. Dealing with such negative values (supposedly proportional to positive optical intensity) is nontrivial, and comes down to the intuition of a data analyst (adding an offset; taking absolute value; or clipping at zero, with no rigorous justification for choosing any of these nonlinear operations over another).

Given such challenges, noise suppression and also feature extraction with no assumption about the Fourier transform of the scattering pattern may become easier with basis functions defined only on pixels with valid intensities and without solely relying on the assumption of the Fourier-Transform-based model.

**Appendix 4: Further aspects of limited-domain basis sets**

*A4.1. Numerical errors*

Despite straightforward *analytical* formulations, there are potential *numerical* issues associated with Bessel function expansions [8]. Including the singular function $Y_\nu$ can make such numerical issues even more considerable. This topic is beyond the scope of this contribution. However, some signatures of such numerical issues can be easily seen and controlled.

An important test parameter is the (Gram) matrix of overlap integrals. If $T$ is the number of the employed orthonormal basis functions $\{\psi_1, \psi_2, \ldots, \psi_T\}$, then the matrix $\boldsymbol{\eta}_{T \times T} = \boldsymbol{\Psi}^H \boldsymbol{\Psi}$ (where $\boldsymbol{\Psi} = [\psi_1 | \psi_2 | \ldots | \psi_T]$) should be ideally the identity matrix $\boldsymbol{\eta}_{T \times T} = \boldsymbol{I}_{T \times T}$. Increasing the smaller radius $a$ or the maximum angular order $\nu_{Max}$, for instance, can result in more pronounced deviations of this matrix from the identity matrix.

Since this issue originates from radial components (Bessel functions), it affects mainly $\langle \psi_{\nu,n} | \psi_{\nu,n'} \rangle$ terms. The signature is non-zero values in the form of blocks (for different permutations of $n$ and $n'$ and a given $\nu$).

More accurate estimations of the roots of the dispersion equation (fine-tuning the roots found with the simple method shown in Appendix 7) is expected to reduce such errors. The role and the significance of Gibbs-like [12] or Pinsky-like [13] issues (with the convergence of a Fourier-like expansion at discontinuities) is to be investigated.

*A4.2. Mappings for other shapes of the ROI boundary*

With perturbations that introduce elongation, but keep the ROI (region of interest) boundary convex, an elliptical approximation may be good enough $x = A\cos(\phi)$ and $y = B\sin(\phi)$

With deformations turning the circular region into a concave curve, a simple conformal mapping [12] as $w = z^n$ (where $z = x + y\sqrt{-1}$ and $w = x' + y'\sqrt{-1}$) is helpful. Special cases of $w = z^2$ and $w = \sqrt{z}$ are classic examples of conformal mapping, which turn a circle into a cardioid and lemniscate, respectively. Rewriting double-integrals in terms of the new coordinate variables requires the calculation of the Jacobian determinant, which is easy thanks to the analytical expression for $w$ as an explicit function of $z$.

*A4.3. Spherical harmonics on a spherical cap*

The scattering pattern generated by a monochromatic beam lies on the *Ewald* surface. In common experiments with small scattering angles, the small spherical cap of detection can be reasonably approximated by the tangential plane at the pole, as implicitly assumed in previous derivations. With large scattering angles, however, the curvature of the domain should be considered.

The starting point for derivation of basis functions was 2D Laplace equation formulated in 2D polar coordinate. A natural generalization is 3D Laplace equation in spherical coordinate and on the surface of a sphere. In the case of the complete coverage of a spherical surface, the eigenfunctions would be spherical harmonics, which are separable in polar and azimuthal variables and parameters as $Y_{l,m}(\theta,\phi) = \Theta_l(\theta)\Phi_m(\phi)$. It makes the reformulation in the case of limited domain easy. Alternative formulation of $\Phi_m(\phi)$ follows the same procedure as addressed in Sec. 3.2. One only needs to evaluate $\Theta_l(\theta)$, subject to new boundaries ($\theta_{min} \leq \theta \leq \theta_{Max}$, as opposed to $0 \leq \theta \leq \pi$) and associated boundary conditions.

The polar function can be written as $\Theta_l(\theta) = c_1 Q_l^m(\cos(\theta)) + c_2 S_l^m(\cos(\theta))$, where $c_1$ and $c_2$ are constants, and $Q_l^m(x)$ and $S_l^m(x)$ are two independent solutions of *general Legendre equation* $[(1-x^2)y']' + [l(l+1) - m^2/(1-x^2)]y = 0$. With full coverage of sphere, non-zero non-singular solutions over the interval $[-1,1]$ are *associated Legendre Polynomials*, denoted by $P_l^m$ and corresponding to $0 \leq l, -l \leq m \leq l$. With a spherical cap, associated Legendre Polynomials $P_l^m$ are replaced with the more general solutions determined within two constants. These two constants are uniquely specified by specific boundary condition at $\cos(\theta_{min})$ and $\cos(\theta_{Max})$ and/or the constraint of non-singularity at pole $\theta_{min} = 0$.

**Appendix 5: Employed Bessel function identities**

Bessel functions are solutions of the ordinary differential equation $[xy']' + (x - v/x)y = 0$ including the parameter $v$. The general solution defined on $[0, \infty)$ is split into the first-order $J_v(x)$ and the second-order $Y_v(x)$ Bessel functions as $y_v(x) = c_v J_v(x) + d_v Y_v(x)$.

The Bessel spectrum of an arbitrary function defined on $[0, \infty)$ can assume a continuous range of $v$ values. However, restricting the domain of definition from $[0, \infty)$ to $[a, b]$ turns the Bessel spectrum into a discrete one written as $\sum_n [c_{v,n} J_v(k_{v,n} x) + d_{v,n} Y_v(k_{v,n} x)]$. A Bessel decomposition defined on a region including the origin will be written as $\sum_n c_{v,n} J_v(k_{v,n} x)$, as $Y_v$ is singular at origin.

Let $v$ be restricted to non-negative integers (denoted by $m$). Also let $C_m$ and $D_m$ represent *either of* the functions $J_m$ or $Y_m$. The following identities are helpful and have been used in the derivations of this contribution:

$$C'_m(z) = C_{m-1}(z) - \frac{m}{z} C_m(z)$$

$$C_{m-2}(x) = \frac{2(m-1)}{x} C_{m-1}(x) - C_m(x)$$

$$qC_m(px)D'_m(qx) - pC'_m(px)D_m(qx) =$$
$$qC_m(px)\left[D_{m-1}(qx) - \frac{m}{qx}D_m(qx)\right] - pD_m(qx)\left[C_{m-1}(px) - \frac{m}{px}C_m(px)\right] =$$
$$qC_m(px)D_{m-1}(qx) - pD_m(qx)C_{m-1}(px)$$

$$\int_0^z C_m(kx)D_m(lx)x\,dx = \frac{x}{k^2 - l^2}[kD_m(lz)C_{m+1}(kz) - lD_{m+1}(lz)C_m(kz)]$$

This overlap integral [14] is valid when the Bessel function index $m$ is a non-negative integer.

**Appendix 6: Overlap integral of orthogonal radial modes**

$$\langle R_{m,n}|R_{m,l}\rangle = \int_{r=a}^{r=b} R_{m,n}(r)R_{m,l}(r)\,rdr$$

$$= \int_{r=a}^{r=b}\left[J_m(k_{m,n}r) - \frac{J_m(k_{m,n}b)}{Y_m(k_{m,n}b)}Y_m(k_{m,n}r)\right]\left[J_m(k_{m,l}r) - \frac{J_m(k_{m,l}b)}{Y_m(k_{m,l}b)}Y_m(k_{m,l}r)\right]rdr$$

$$= \int_{r=a}^{r=b} J_m(k_{m,n}r)J_m(k_{m,l}r)rdr$$

$$-\frac{J_m(k_{m,l}b)}{Y_m(k_{m,l}b)}\int_{r=a}^{r=b} J_m(k_{m,n}r)Y_m(k_{m,l}r)rdr$$

$$-\frac{J_m(k_{m,n}b)}{Y_m(k_{m,n}b)}\int_{r=a}^{r=b} Y_m(k_{m,n}r)J_m(k_{m,l}r)rdr$$

$$+\frac{J_m(k_{m,n}b)}{Y_m(k_{m,n}b)}\frac{J_m(k_{m,l}b)}{Y_m(k_{m,l}b)}\int_{r=a}^{r=b} Y_m(k_{m,n}r)Y_m(k_{m,l}r)rdr$$

If $C$ and $D$ denote any Bessel function (either $J$ or $Y$), the following formula holds for arbitrary positive argument $x$ and non-negative integer parameter $m$:

$$\psi_{p,q}^{C,D}(m,x) = \int_{r=0}^{r=x} C_m(pr)D_m(qr)rdr$$

$$= \frac{x}{p^2-q^2}[qC_m(px)D_{m+1}(qx) - pC_{m+1}(px)D_m(qx)]$$

For the useful degenerate case, we use L'Hopital's rule to derive

$$\psi_{q,q}^{C,D}(m,x) = \lim_{p\to q}\psi_{p,q}^{C,D}(m,x)$$

$$= \frac{x}{2q}[qxC_m'(qx)D_{m+1}(qx) - D_m(qx)[C_{m+1}(qx) + xqC_{m+1}'(qx)]] =$$

$$\frac{x}{2q}\left[qxD_{m+1}(qx)\left[C_{m-1}(qx) - \frac{m}{qx}C_m(qx)\right] - D_m(qx)C_{m+1}(qx)\right.$$

$$\left. - D_m(qx)xq\left[C_m(qx) - \frac{m-1}{qx}C_{m+1}(qx)\right]\right]$$

After simplification

$$\psi_{q,q}^{C,D}(m,x) = \frac{x}{2q}D_{m+1}(qx)(qxC_{m-1}(qx) - mC_m(qx))$$

$$+ \frac{x}{2q}D_m(qx)(-C_{m+1}(qx) + (m-2)qxC_m(qx))$$

The overlap integral $\langle R_{m,n}|R_{m,n}\rangle$ requires subtraction of the indefinite integral $\psi_{q,q}^{C,D}(m,x)$ at boundaries. Defining definite integrals as $\phi_n^{C,D}(m,b,a) \equiv \psi_{k_{m,n},k_{m,n}}^{C,D}(m,b) - \psi_{k_{m,n},k_{m,n}}^{C,D}(m,a)$, one can write

$$\langle R_{m,n}|R_{m,n}\rangle = \phi_n^{J,J}(m,b,a) - 2\left[\frac{J_m(k_{m,n}b)}{Y_m(k_{m,n}b)}\right]\phi_n^{J,Y}(m,b,a) + \left[\frac{J_m(k_{m,n}b)}{Y_m(k_{m,n}b)}\right]^2\phi_n^{Y,Y}(m,b,a)$$

**Appendix 7: Matlab implementation of the key formulae**

*A7.1. Dispersion equation*
```
%Input m: Order of Bessel function
%Input a: Smaller radius defining the region of interest (ROI)
%Input b: Larger radius defining the region of interest (ROI)
%Input n_max: Maximum number of roots for a given m
%Output k_m_n: Roots of Dispersion Equation
%Approach: Global search and looking for sign changes
function k_m_n=Dispersion(m,a,b,n_Max)
    N_scan=1e5;
    d=a/b;
    x=linspace(0,8*n_Max,N_scan);
    Threshold1=1e-1;
    Threshold2=2*(x(2)-x(1));

f=besselj(m,x)./bessely(m,x)+BesselD(m,d*x,'J')./BesselD(m,d*x,'Y');
    x_=x(2:(end-1));
    g=f(1:(end-2)).*f(3:end);
    Index=(g < 0) & (abs(f(2:(end-1))) < Threshold1);
    xRoots=x_(Index);
    for cntr=2:numel(xRoots)
        if ~isnan(xRoots(cntr-1))
            if abs(xRoots(cntr)-xRoots(cntr-1)) < Threshold2
                xRoots(cntr)=nan;
            end
        end
    end
    xRoots=xRoots(~isnan(xRoots));
    xRoots=xRoots(1:n_Max);
    k_m_n=xRoots/b;
    Plot_Flag=1;
    if Plot_Flag
        H=figure;
        plot(x,f,'b.-',xRoots,xRoots*0,'r*');
        title(num2str(sum(xRoots)))
        ylim([-1 1]);
        drawnow;
        pause;
        close(H);drawnow;
    end
end
```

*A7.2. Derivatives of Bessel functions*
```
%Input m: Order of Bessel function
%Input x: Argument of Bessel function
%Input Type: Type of Bessel function (string)
%Output Bd: Analytical derivative of the Bessel function
function Bd=BesselD(m,x,Type)
    switch Type
        case 'J'
```

```
            Bd=besselj(m-1,x)-(m./x).*besselj(m,x);
        case 'Y'
            Bd=bessely(m-1,x)-(m./x).*bessely(m,x);
    end
end
```

*A7.3. Kernel matrix of 2D Discrete Fourier Transform*
```
%Input n: Number of pixels per coordinate
%Output G: The Kernel matrix
N=n^2;
G=rand(N,N);
Exp=exp(-1i*(2*pi/n));
WB=waitbar(0,'Calculating the Kernel matrix');drawnow;
for r=0:(n-1)
    for s=0:(n-1)
        Lf=r*n+s;
        for p=0:(n-1)
            for q=0:(n-1)
                LF=p*n+q;
                G(LF+1,Lf+1)=Exp.^(r*p+s*q);
            End
            %Vector-based coding possible with 1D or 2D grid
            %Alternative 1D Vector-based coding:
            %Q=(0:(n-1))';
            %G(p*n+1+Q,Lf+1)=Exp.^(r*p+s*Q);
        end
        WB=waitbar(Lf/N,WB);drawnow;
    end
end
close(WB);drawnow;
```

*A7.4. Linear retrieval of (non-trivial) Patterson profile*
```
%Input n: Number of pixels per coordinate
%Input S: Indices of nontrivial Patterson values (in Support)
%Input K: Indices of known pixels
%Input G: The Kernel matrix, as calculated before
%Input F: The FULL scattering pattern with swapped quadrants
%Output Patterson: The FULL Patterson profile
%Convention for 1D <-> 2D indexing: As in Matlab; a_1D=a_2D(:)
Threshold=1e-12;
Patterson=zeros(n,n);
Patterson(S)=real(lsqr(G(K,S),F(K),Threshold));
```